\documentclass[manuscript]{acmart}

\usepackage[acronym,nonumberlist]{glossaries}
\glsdisablehyper
\graphicspath{{./}}

\newacronym{LPA}{LPA}{latent profile analysis}
\newacronym{BFI-10}{BFI-10}{10-item Big Five Inventory}
\newacronym{CRT}{CRT}{Cognitive Reflection Test}
\newacronym{BIC}{BIC}{Bayesian information criterion}
\newacronym{AIC}{AIC}{Akaike information criterion}
\newacronym{BLRT}{BLRT}{bootstrap likelihood-ratio test}
\newacronym{HSD}{HSD}{honestly significant difference}
\newacronym{CI}{CI}{confidence interval}
\newacronym{OR}{OR}{odds ratio}
\newacronym{SE}{SE}{standard error}
\newacronym{AI}{AI}{artificial intelligence}
\newacronym{JITAI}{JITAI}{just-in-time adaptive intervention}
\newglossaryentry{IOS}{name={iOS},description={Apple's mobile operating system}}
\newacronym{ANOVA}{ANOVA}{analysis of variance}

\newcommand{\appname}{Structured}
\newcommand{\appurl}{\texttt{structured.app}}

\renewcommand\footnotetextcopyrightpermission[1]{}

\title{How do people plan digitally: An in-the-wild investigation of task planning through a smartphone app}

\author{Srija Halder}
\email{shalder@clarku.edu}
\affiliation{%
  \institution{Clark University}
  \city{Worcester}
  \state{MA}
  \country{USA}
}

\author{Isabella Zimmermann}
\affiliation{%
  \institution{Structured}
  \city{Berlin}
  \country{Germany}
}

\author{Linnea K\"orte}
\affiliation{%
  \institution{Structured}
  \city{Berlin}
  \country{Germany}
}

\author{Leo Mehlig}
\affiliation{%
  \institution{Structured}
  \city{Berlin}
  \country{Germany}
}

\author{Paul Schmiedmayer}
\affiliation{%
  \institution{Stanford University}
  \city{Stanford}
  \state{CA}
  \country{USA}
}

\author{David Gr\"uning}
\email{gruening@stanford.edu}
\affiliation{%
  \institution{Stanford University}
  \city{Stanford}
  \state{CA}
  \country{USA}
}
\affiliation{%
  \institution{Max Planck Institute for Human Development}
  \city{Berlin}
  \country{Germany}
}
\affiliation{%
  \institution{University of Cambridge}
  \city{Cambridge}
  \country{UK}
}

\renewcommand{\shortauthors}{Halder et al.}
\authorsaddresses{Corresponding authors: Srija Halder,
  \href{mailto:shalder@clarku.edu}{shalder@clarku.edu}. David Grüning,
  \href{mailto:gruening@stanford.edu}{gruening@stanford.edu}.}

\begin{abstract}
Daily planning supports goal attainment and productivity, and people increasingly delegate it to digital tools. Plans are postponed, revised, and left unresolved rather than executed as intended. Understanding these changes requires following tasks from creation through later updates and recorded outcomes. We report an in-the-wild study of task planning using time-stamped logs of 24,265 tasks from 957 users of a widely used daily planner app, observed over six weeks alongside self-reported surveys. Following each task across its lifecycle, only 26.4\% moved directly from creation to completion and 32.0\% were abandoned; all-day and longer tasks were abandoned disproportionately, and 77.6\% of timed tasks were marked complete later than intended. Latent profile analysis of 909 users indicated High Engagement (11.1\%), Low Engagement (19.7\%), and Passive (69.2\%) profiles; active days on app distinguished the profiles and were associated with post-survey completion. These findings support evaluating planning tools across the task lifecycle.
\end{abstract}



\begin{document}
\maketitle
\glsresetall


\section{Introduction}
Planning is central to how people pursue their goals.
Deciding what to do, when to do it, and in what order lets people convert broad aspirations into concrete, scheduled actions, and the everyday capacity to organize one's time is tied to productivity and, through it, to well being.
Calendars, reminder systems, and dedicated planner applications support this everyday work of organizing tasks and schedules.

Their interaction model, however, is built around a narrower path than the empirical literature describes: a task is entered once, carries a single scheduled time, and is closed by being marked complete. Deferral, revision, and quiet abandonment are representable only as the absence of that final act. We call this the \emph{plan-and-execute} view, and we mean it as a description of how planning tools are structured and evaluated, not as a claim that researchers believe plans are always followed.
Under it, the psychologically interesting moment is the formation of an intention; what follows is treated as routine implementation.
On this account, a good planning tool is essentially a faithful ledger of intentions, and a good plan is one that is carried out as written.

Decades of research on self-regulation suggest that this assumption is the exception rather than the rule.
Intentions frequently fail to translate into action, the well-documented intention--behavior gap~\cite{sheeran2016intention, webb2006does}.
Procrastination, described as a quintessential form of self-regulatory failure, is both prevalent and systematically predicted by task aversiveness, temporal delay, impulsiveness, and low conscientiousness~\cite{steel2007nature}.
The reliable effectiveness of remedies such as implementation intentions~\cite{gollwitzer1999implementation, gollwitzer2006implementation} is itself evidence that divergence is the default: if plans were routinely executed as formed, there would be little for such strategies to repair.
Plans can be postponed, modified, or abandoned after an intention is formed.

If plans routinely change after they are made, then understanding planning requires observing what happens to a task across its entire lifecycle: not merely whether an intention was formed, but how that intention is revised, deferred, or dropped as time passes.
This shifts the object of study from a single decision to a temporally extended process, and it raises questions that a plan-and-execute framing cannot answer: How often are tasks rescheduled before they are completed?
How long do tasks survive before they are abandoned?
When, in the rhythm of a day or a week, does planning actually happen?
And do these dynamics differ systematically across people?

We address these questions with an in-the-wild investigation of task planning using naturalistic usage logs from a widely used daily planner application (\appname; \appurl), a mobile app that offers a detailed, adaptive interface for semi-guided, digitally assisted planning, complemented by surveys of personality, cognition, and self-perceived productivity and wellbeing.
The logs record time-stamped task events, allowing us to follow in-app planning over time and match these records to survey responses.

The central contribution of this work is empirical: a large corpus of naturalistic, time-stamped planning behavior from real users planning their real lives.
These records let us follow task updates and recorded outcomes without asking users to reconstruct them retrospectively.
Our primary analyses therefore follow the task across its lifecycle.
Using creation, scheduling, rescheduling, completion, and abandonment timestamps, we trace how tasks move from intention to outcome: how many reach completion, how long they survive before they are finished or abandoned, how often they are deferred along the way, and when in the day and week planning and completion actually occur.
We relate these dynamics to task characteristics, estimated duration, subtask structure, and temporal specificity (all-day versus timed), and to individual differences captured through personality and cognition surveys.

Building on this lifecycle account, we then ask whether the ways people plan cluster into distinct types.
We derive data-driven user profiles from core in-app behavior and characterize how they differ in their activity over time and their likelihood of dropout, interpreting them psychologically using self-report.
The profiles are a lens on the lifecycle, not a substitute for it: they describe differences in app use across users, but the object of study throughout is what happens to plans after they are made.
The two analyses describe task histories and differences in app use across users.

This paper makes three contributions.
\emph{First}, an empirical one: a task-level account of the planning lifecycle in the wild, tracing 24{,}265 tasks from 957 users across creation, rescheduling, and completion or abandonment, with time-to-event estimates, rescheduling counts, and the diurnal and weekly rhythm of planning.
\emph{Second}, a conceptual one: an account of task revisions and unresolved plans that complements measures of creation and completion.
\emph{Third}, a design-relevant one: a behavioral typology of app use and an analysis linking active days to post-survey response, informing hypotheses for continued engagement.

\section{Related Work}

\subsection{Task management and planning in the wild}
Naturalistic studies of task management have long recognized that to-dos are recorded across many media and completed unpredictably over time~\cite{bellotti2004todo}.
That early work established that a personal task list is rarely a tidy queue of items dispatched in order; instead, tasks are captured opportunistically, revisited, and frequently left unresolved.
Much of what is known about planning and its failures, however, still comes from laboratory experiments using artificial tasks and from self-report and retrospective accounts.
Each of these has well-known limitations.
Laboratory tasks strip away the competing demands, interruptions, and stakes that characterize real planning; retrospection and self-report are subject to memory distortion and social-desirability bias, so that reported behavior may differ substantially from actual behavior.
Large-scale behavioral traces of everyday planning, data that capture what people do, moment to moment, remain comparatively scarce.
What is missing is not another account of why intentions fail, but a record of planning dense enough to watch it happen: the moment-to-moment trace of tasks being created, deferred, and resolved or dropped in everyday life.
The most basic gap is therefore observational, the full task lifecycle, from creation through rescheduling to completion or abandonment, has rarely been watched end to end in the wild, and, bound up with it, temporal: when planning occurs and how delays accumulate and resolve over hours, days, and weeks.
Only once planning can be observed this way does the question of whether these dynamics differ systematically across people become tractable.
This study is built to close the observational and temporal gaps directly and to make the individual-differences question approachable.

\subsection{Behavioral logs and individual differences}
A growing body of work uses passively collected smartphone behavior to characterize people rather than only their momentary actions.
Patterns of everyday phone behavior can predict stable personality characteristics~\cite{stachl2020predicting}, motivating the more general strategy of inferring psychological types from unobtrusive logs.
Within digital health and behavior-change systems, latent profile analysis has been used to recover interpretable subgroups of users from engagement data~\cite{choi2023patterns}, and platform activity itself has been shown to moderate how effective an intervention is~\cite{katsaros2024platform}.
We build on this tradition by applying profile analysis to the specific behaviors of digital planning, task creation, completion, recurrence, and the temporal breadth of engagement, and by linking the recovered profiles to self-report.

\subsection{Self-regulation and the intention--behavior gap}
Theories of goal pursuit provide the conceptual backdrop for why plans fail.
The theory of planned behavior treats intention as the proximal cause of action~\cite{ajzen1991theory}, yet meta-analyses show that changing intentions produces only modest changes in behavior~\cite{webb2006does, sheeran2016intention}, and procrastination is a robust, trait-linked form of self-regulatory failure~\cite{steel2007nature}.
Implementation intentions, specific if-then plans that link a situational cue to a goal-directed response, are among the most reliable remedies~\cite{gollwitzer1999implementation, gollwitzer2006implementation}.
The existence and efficacy of such remedies is precisely why a lifecycle view is warranted: it presupposes that intentions, once formed, routinely diverge from action.

\subsection{Digital tools for productivity and behavior change}
A large design literature seeks to close this gap with technology.
Reminder systems and electronic prompts improve adherence in health contexts~\cite{vervloet2012effectiveness}, and mobile systems that add implementation-intention prompts or plan reminders can support behavior change~\cite{pirolli2017implementation, wicaksono2019investigating}.
Digital care and workplace-wellbeing programs report effects on productivity and mood~\cite{costa2022impacts, kulkarni2022impact}, though sustained engagement is a recurring obstacle: friction-based and other interventions tend to see use decline rapidly after installation~\cite{haliburton2024understanding}.
Two framings are particularly relevant to how a planner might help.
\Gls{JITAI} frameworks argue that support should be delivered in the right form at the right moment~\cite{nahumshani2018jitai}, and the self-nudging / boosting perspective casts the user as a choice architect who reshapes their own environment with the tool's help~\cite{reijula2022selfnudging, hertwig2017boosting}.
Our profile-aware design implications draw on both.

\section{Method}
We combined two complementary streams of field data.
First, we passively logged users' in-app planning behavior across a six-week observation window, yielding time-stamped records of the tasks they created, scheduled, rescheduled, and completed.
Second, we administered brief self-report surveys immediately before and after this window, capturing personality, cognitive reflection, related social-cognitive tendencies, and self-perceived productivity and wellbeing.
The two streams link in-app planning records from new users of \appname{} to their survey responses.

\subsection{Data source and application}
Data come from usage logs of \appname{} (\appurl), a commercially available daily planner application for \gls{IOS} and Android in which users create, schedule, edit, and complete tasks.
\appname{} provides a detailed and adaptive interface for semi-guided, digitally assisted planning: users assemble a structured schedule of their planned tasks and can annotate each scheduled item with additional motivating or specifying information.
Relevant to the lifecycle measures used here, the app time-stamps the creation of each task, lets users schedule it as an all-day or timed item for a specific day and time, supports later editing and rescheduling, allows tasks to recur, and registers whether the task is ultimately marked as completed.

\subsection{Participants and sampling}
New, anonymous, and voluntary \gls{IOS} users of \appname{} were recruited through an in-app pop-up message inviting them to take part in a study about the app.
Interested users were forwarded to the baseline survey and provided informed consent before any data were recorded.
In total, $N = 1{,}932$ new users were recruited, and their behavior was observed from the onset of use.
Each participant's in-app activity was then monitored for six weeks, and if a participant deleted the app or ended the study early, data were retained only up to that point.
To limit the identifying information collected, we did not collect sociodemographic information such as age or sex.

Two analytic samples follow from this design, and we distinguish them throughout.
The \emph{lifecycle sample} comprises the $N = 957$ recruited users who generated at least one usable task record (creation, completion, incompletion or update), contributing 24,265 tasks; it underpins all task-level analyses, which require only logs.
The \emph{profile sample} comprises the $N = 909$ users who created at least one new task during the observation window, enabling computation of the behavioral indicators underpinning the latent profile analysis. Of these, $N = 904$ were retained for analyses linking profiles to survey outcomes. From the 904, between 70 and 74 users (7.7--8.2\%) had completed the follow-up survey and contribute to the outcome analyses; the exact sample size for each outcome is reported in the results.

\subsection{Behavioral (log) measures}

Lifecycle events were derived from time-stamped log entries.
For each task, the app recorded its creation, its intended completion time, any rescheduling of the item, and whether it was ultimately completed or left incomplete, together with the associated timestamps; the analyses use these events to derive task histories.
From the logs we constructed the four core behavioral indicators used in the profile analyses: the number of tasks a user created (log-transformed), the proportion of created tasks that were completed (completion rate), the proportion of tasks set up as recurring items (recurring-task rate), and the number of distinct calendar days on which the user was active in the app (active days).
Temporal specificity was captured as all-day versus timed scheduling.
Dropout was operationalized in two ways: behaviorally, as failure to complete the six-week observation period (deleting the app or terminating early), and, for the survey component, as non-completion of the post survey after the study weeks.

\subsection{Survey measures}
Before and after the study, users completed brief self-reports.
Specifcially, we measured their general wellbeing (1 - very unhappy to 7 - very happy), productivity (percentage of indicated tasks actually completed in the day), happiness with their productivity (1 - very unhappy to 7 - ver happy), and how problematic they found their productivity (1 - very unproblematic to 7 - very problematic).
Personality was assessed with the \gls{BFI-10}~\cite{rammstedt2014bfi}, covering extraversion, agreeableness, conscientiousness, neuroticism, and openness.
Cognition was assessed with the \gls{CRT} (scored 0--7)~\cite{frederick2005cognitive}, which captures the tendency to override an intuitive but incorrect response with a reflective, correct one.

We additionally measured two social-cognitive tendencies adapted to the productivity domain.
Self-enhancement, a productivity analogue of the better-than-average effect, was measured by asking users to rate their own productivity relative to the average person from 1 (very below average) to 7 (very above average).
Social projection was operationalized as the difference between a user's absolute rating of their own productivity and their rating of people's productivity in general (both on 1--7 scales), with smaller differences indicating stronger projection.
Self-perceived productivity problems were indexed by the absolute self-rating of one's own (un)productiveness, and wellbeing by the general wellbeing item.
Item wording paralleled that of \citet{gruning2023social}, with content shifted from problematic digital consumption to productivity.

Only measured in the post survey were participants' perceived progress with their goals through using the app (1 - not at all to 5 - completely reached the goals) and their satisfaction with their progress due to the app (1 - very unsatisfied to 7 - very satsified). A general wellbeing item (1 - very unhappy to 7 - very happy) was also administered once per week during the study time to capture wellbeing-experiences.

\subsection{Analytic approach}
Descriptive analyses summarized lifecycle events, task characteristics, and their temporal dynamics.
To identify types of use, the four core behavioral indicators, tasks created (log-transformed), completion rate, recurring-task rate, and active days, were submitted to \gls{LPA} using the \texttt{tidyLPA} package~\cite{rosenberg2018tidylpa} in R.
Solutions with two to seven profiles were estimated, and the retained solution was chosen by weighing the \gls{BIC} and \gls{AIC}, classification certainty (entropy), the \gls{BLRT}, the size of the smallest profile, and theoretical interpretability.
Differences between profiles in subjective outcomes were tested with one-way analyses of variance, with Tukey \gls{HSD} comparisons correcting for multiple post-hoc contrasts; differences in post-survey completion were tested with a chi-square test; and dropout was modeled with logistic regression predicting post-survey completion from the behavioral indicators.

To characterize the task lifecycle directly, we additionally analyzed the time-stamped event history of individual tasks.
We summarized each task's fate as a set of state transitions (created $\rightarrow$ scheduled $\rightarrow$ rescheduled $\rightarrow$ completed or abandoned) and estimated the cumulative incidence of completion and of abandonment over time using competing-risks survival analysis, treating tasks still open at the end of the six-week window as censored.
Rescheduling was analyzed as an event count per task and related to eventual completion.
We described the time-of-day and day-of-week rhythm of planning using circular summaries, and we quantified the lag between a task's intended completion time and its actual completion.
Finally, we related task characteristics, estimated duration, subtask structure, and temporal specificity (all-day versus timed), to lifecycle outcomes.

\subsection{Ethics and data protection}
The study was assessed internally at Heidelberg University, which determined that it did not require a formal ethics vote, on the grounds that the data collected are neither sensitive nor suitable for identifying individuals.

All participants were adults and gave informed consent.
Recruitment was by an in-app invitation shown to new users, which could be dismissed without consequence; users who expressed interest were forwarded to the baseline survey, and no behavioral data were recorded until consent had been given there.
Participation was voluntary and unpaid, and participants could withdraw at any point by ending the study early or deleting the app, in which case logging ceased and only data generated up to that point were retained.

Data minimization was part of the design rather than a subsequent step.
No sociodemographic information was collected; this limits the information available for identifying individuals but does not establish that linked behavioral records cannot be re-identified.
Logs of everyday task management can be revealing even without task content.
The corpus is not released in raw form; the terms of access are described below.

\subsection{Accessibility of data and code}
Analysis code and the derived, de-identified data supporting the reported analyses are available in the project's Open Science Framework repository (\url{https://osf.io/3m8vj/}), together with a codebook for the behavioral indicators and the survey instruments.
Neither the analyses nor the study procedures were preregistered.
The raw event logs are not deposited: they are held under the terms governing the application's user data, and are available from the authors on reasonable request subject to those terms.

\section{Results}

\subsection{The task lifecycle}
Before turning to person-level profiles, we describe what happens to tasks themselves across their lifecycle; these descriptive dynamics are the primary object of the present study, and the user profiles reported subsequently build on them.
Analyses in this subsection use the lifecycle sample of 24,265 tasks from 957 users.

\begin{table*}[t]
\caption{Task fate by task characteristic, and estimated duration by fate.
The three fates are mutually exclusive and exhaustive.}
  \label{tab:task-fate}
  \centering
  \begin{tabular}{lrrrrr}
    \toprule
    \textbf{Characteristic} & \textbf{$n$} & \textbf{\% Completed} & \textbf{\% Abandoned} & \textbf{\% Still open} & \textbf{$\chi^2(2)$}\\
    \midrule
    \multicolumn{6}{l}{\textit{Temporal specificity}}\\
    \quad Timed tasks    & 23{,}848 & 49.9 & 31.5 & 18.6 & \\
    \quad All-day tasks  &      416 & 23.1 & 61.8 & 15.1 & 179.39\textsuperscript{***}\\
    \addlinespace
    \multicolumn{6}{l}{\textit{Subtask structure}}\\
    \quad No subtasks    & 23{,}295 & 49.6 & 32.3 & 18.0 & \\
    \quad Has subtasks   &      970 & 45.8 & 23.9 & 30.3 & 99.43\textsuperscript{***}\\
    \midrule
    \textbf{Fate} & \textbf{$n$} & \textbf{$M$ duration} & \textbf{$Mdn$ duration} & & \\
    \midrule
    \quad Completed & 12{,}006 & 60.8 min & 30 min & & \\
    \quad Abandoned &  7{,}767 & 75.9 min & 45 min & & \\
    \quad Still open &  4{,}492 & 45 min   & 45 min & & \\
    \bottomrule
  \end{tabular}

  \vspace{3pt}
  \begin{minipage}{\textwidth}
    \footnotesize
    \textit{Note.} Upper panel: $\chi^2$ tests compare the distribution of task fate across characteristic groups ($\mathit{df} = 2$).
    ``Still open'' denotes tasks that had reached no terminal state by the end of the six-week window.
    Lower panel: estimated duration is the user-entered time estimate attached to a task, reported only for tasks carrying one.
    The association between estimated duration and completion was tested via logistic regression: $\glsentryshort{OR} = 0.999$ per minute, 95\% \gls{CI} $[0.999, 0.999]$, $p < .001$.
    \textsuperscript{***}$p < .001$.
  \end{minipage}
\Description{Two-panel table.
The upper panel breaks task fate (percent completed, abandoned, and still open) down by temporal specificity (timed versus all-day tasks) and by subtask structure (with versus without subtasks), with chi-square tests; all-day tasks and tasks with subtasks show markedly different fate distributions from their counterparts, both significant at p less than .001.
The lower panel gives mean and median user-estimated duration for completed, abandoned, and still-open tasks, with abandoned tasks longer than completed ones.}
\end{table*}

\subsubsection{From creation to completion or abandonment}
\begin{figure}[t]
  \centering
  \includegraphics[width=\columnwidth]{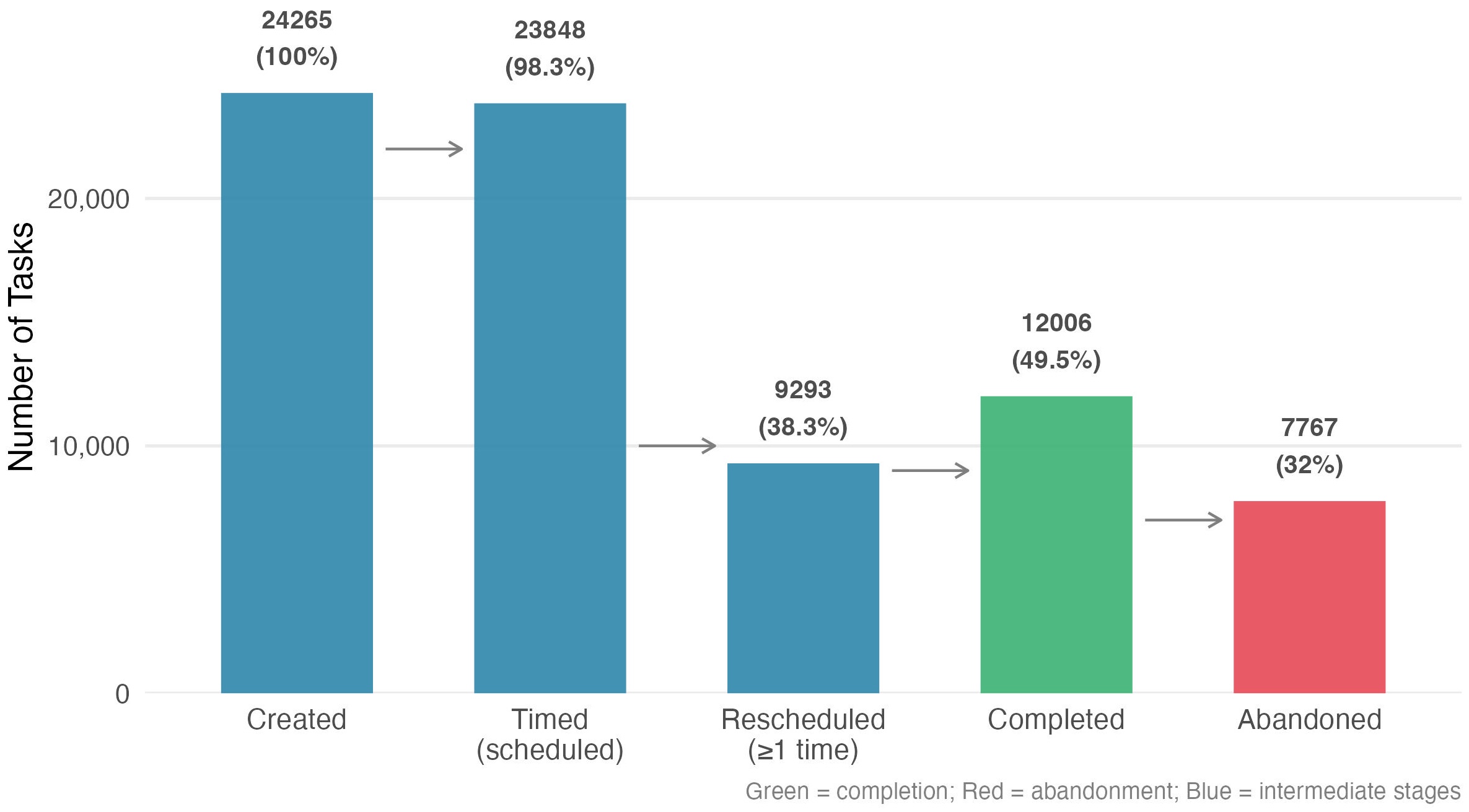}
  \caption{Task lifecycle as a state-transition funnel.}
  \Description{Funnel showing proportions of tasks moving from creation
    to scheduling, rescheduling, and terminal completion or abandonment.}
  \label{fig:funnel}
  \vspace{2pt}
  \parbox{\columnwidth}{\footnotesize \textit{Note.} State-transition funnel across 24,265 tasks from 957 users.}
\end{figure}

Across the 24,265 tasks created by the 957 users, 49.5\% ($n = 12{,}006$) were ultimately marked complete, 32.0\% ($n = 7{,}767$) were effectively abandoned (created but neither completed nor carried forward), and 18.5\% ($n = 4{,}492$) remained open at the end of the observation window.
Figure~\ref{fig:funnel} presents the task lifecycle as a state-transition funnel, showing the proportion of tasks moving from creation to scheduling, rescheduling, and terminal completion or abandonment.
Contrary to the plan-and-execute assumption, only a minority of created tasks (26.4\%) followed the direct creation-to-completion path; the remaining 23.1\% of completions arrived only after at least one reschedule, and about half had no recorded completion within the observation window.

\subsubsection{Time to completion and time to abandonment}
\begin{figure}[t]
  \centering
  \includegraphics[width=\columnwidth]{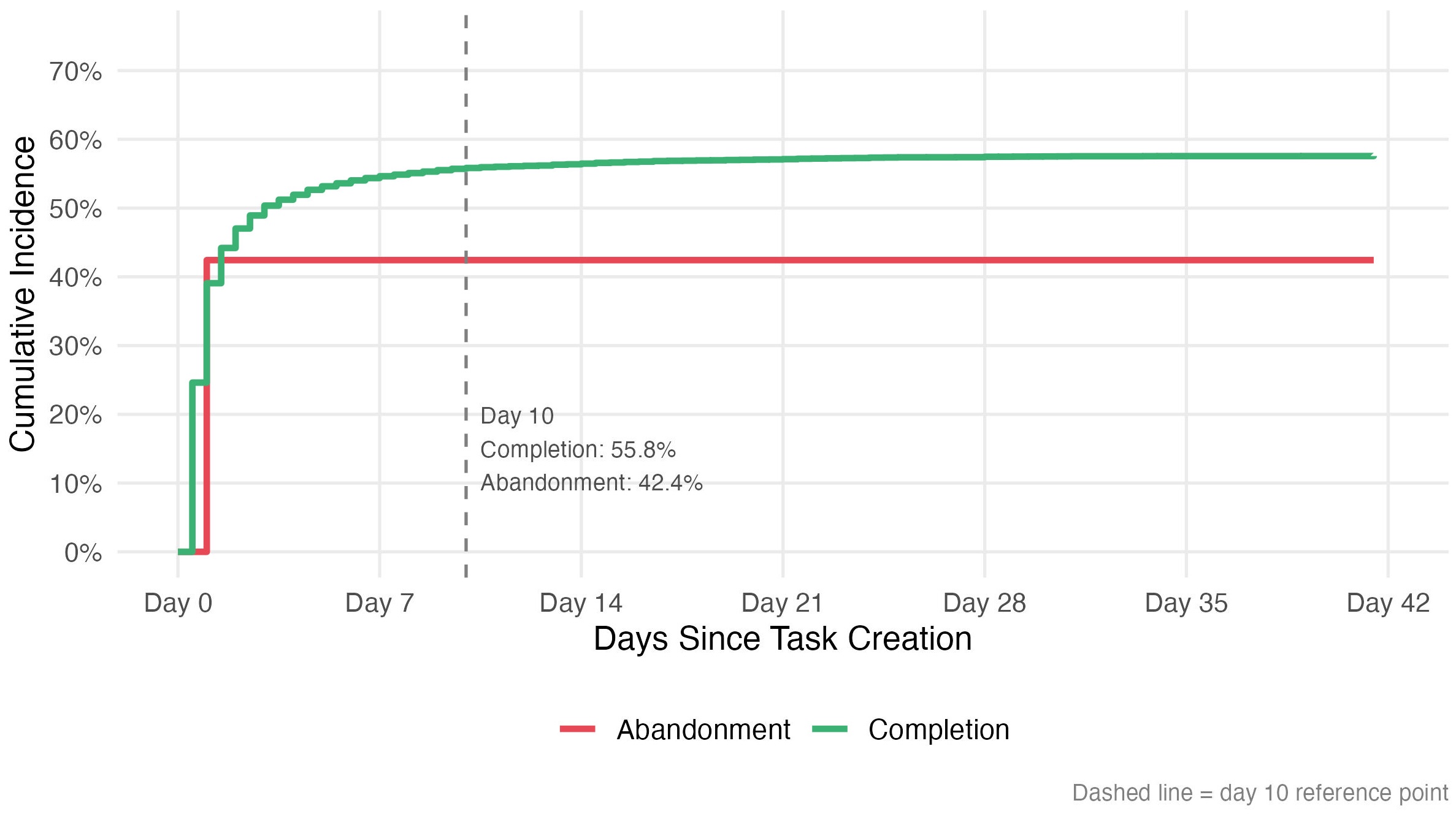}
  \caption{Cumulative incidence of task completion and abandonment.}
  \Description{Cumulative Incidence of Task Completion and Abandonment.}
  \label{fig:risks}
  \vspace{2pt}
  \parbox{\columnwidth}{\footnotesize \textit{Note.} Competing-risks analysis; tasks still open at window end censored.}
\end{figure}

We estimated the cumulative incidence of completion and of abandonment as a function of time since task creation, treating the two outcomes as competing risks and censoring tasks still open at the window end (Figure~\ref{fig:risks}).
Completion was fast when it happened at all: completed tasks were resolved a median of 0.50 days after creation (mean 1.44 days), indicating a short interval between creation and recorded completion for many completed tasks.
Abandonment accrued more slowly, with cumulative incidence reaching 42.4\% by day 10 and remaining essentially flat thereafter, with little additional cumulative incidence of abandonment after day 10 in this observation window.
At day 10 the cumulative incidence of completion stood at 55.8\% against 42.4\% for abandonment, a ratio of 1.32.
These censoring-adjusted incidences exceed the crude proportions reported above (49.5\% and 32.0\%) as tasks created late in the observation window contribute only partial follow-up; the cumulative incidences also depend on the censoring assumptions.

\subsubsection{Rescheduling}

Most tasks were never deferred, but a sizeable minority were deferred repeatedly.
A majority of tasks (58.4\%) were never rescheduled, while 38.3\% were rescheduled three or more times; the distribution is accordingly heavily right-skewed (median 0, mean 12.93, 90th percentile 36, maximum 1,105).
Reschedule frequency was associated with eventual fate, but not in the direction a simple procrastination account predicts: each additional reschedule corresponded to a slight \emph{increase} in the odds of completion ($\glsentryshort{OR} = 1.007$, 95\% \gls{CI} $[1.007, 1.008]$).
We read this as survivorship rather than benefit.
A task that is still being moved is a task the user has not yet written off, so repeated deferral marks a plan that remains live, whereas abandonment tends to be silent; the task is simply never touched again.
Deferral and abandonment are thus distinct fates rather than points on one continuum, which is precisely the distinction a plan-and-execute framing collapses.

Two features of this distribution warrant caution.
The extreme upper tail is unlikely to represent deliberate acts of re-planning: values in the hundreds are more plausibly generated by recurring items, whose instances the app rolls forward automatically, than by a user repeatedly choosing a new date.
The mean is correspondingly uninformative about typical behavior, and we rely on the quantiles instead.
Reschedule count also accumulates while a task remains open and therefore depends partly on exposure time.
The odds ratio is descriptive and does not establish a benefit of rescheduling.

\subsubsection{When planning happens}
\begin{figure}[t]
  \centering
  \includegraphics[width=\columnwidth]{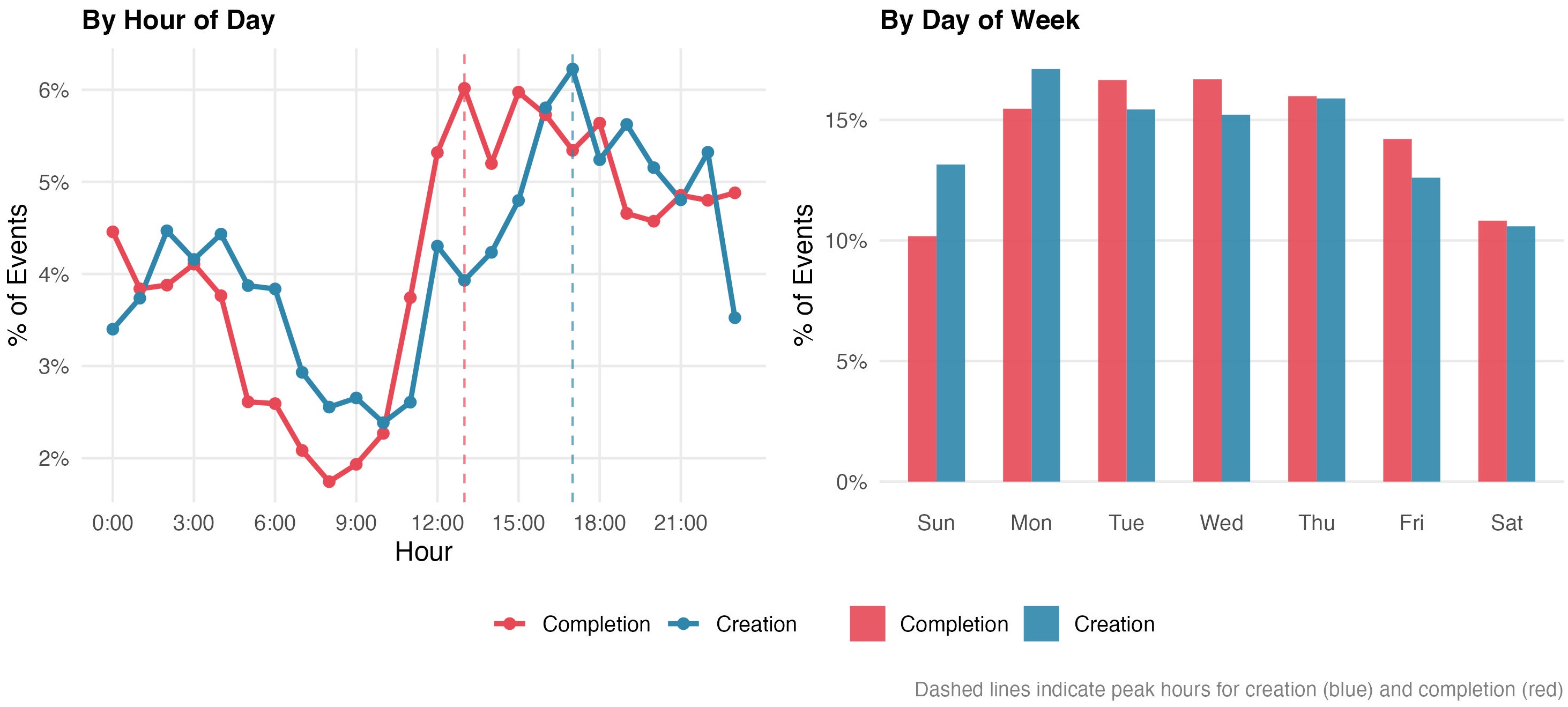}
  \caption{Temporal distribution of task creation and completion.}
  \Description{Temporal distribution of task creation and completion.}
  \label{fig:distr}
  \vspace{2pt}
  \parbox{\columnwidth}{\footnotesize \textit{Note.} Creation peaks at 5:00 PM on Mondays; completion peaks at 1:00 PM on Wednesdays.}
\end{figure}

Task creation and completion were unevenly distributed across the day and week (Figure~\ref{fig:distr}).
Creation peaked in the evening (5:00 PM) on Monday, while completions clustered at midday (1:00 PM) on Wednesday.
The typical lag between a task's intended completion time and its recorded completion was 1 hour (mean 3.9 hours); 77.6\% of timed tasks were marked complete later than intended.

\subsubsection{Task characteristics and their fate}

Task fate varied systematically with how a task was specified (Table~\ref{tab:task-fate}).
Temporal specificity mattered most.
All-day tasks were disproportionately abandoned (61.8\% vs.\ 31.5\% for timed tasks) and correspondingly less often completed (23.1\% vs.\ 49.9\%), $\chi^2(2) = 179.39$, $p < .001$.
Committing to a time, rather than merely to a day, thus accompanied a roughly twofold difference in the likelihood of a task being carried out, the pattern implementation-intention theory would predict, observed here in ordinary use rather than under instruction.
Size mattered too, though less sharply: abandoned tasks carried longer duration estimates than completed ones (means 75.9 vs.\ 60.8 min; medians 45 vs.\ 30 min), and each additional estimated minute slightly reduced the odds of completion ($\glsentryshort{OR} = 0.999$, 95\% \gls{CI} $[0.999, 0.999]$, $p < .001$).
Subtask structure showed a third pattern: tasks broken into subtasks were less often abandoned (23.9\% vs.\ 32.3\%) but were also the most likely to be left open at the window's end (30.3\% vs.\ 18.0\%), $\chi^2(2) = 99.43$, $p < .001$, as though decomposition kept a task alive without bringing it to a close.
Across all three characteristics, how a task is specified at creation is associated with where it ends up in the lifecycle.

\subsection{Latent profile analysis: model selection}

We estimated \glsentryshort{LPA} solutions with two to seven profiles.
Fit statistics for all models are presented in Table~\ref{tab:fit}.
\glsentryshort{BIC} values decreased across all tested solutions, favoring models with more than three profiles on this criterion.
Entropy ranged from 0.876 to 0.900 across solutions.
The \glsentryshort{BLRT} was significant for all solutions tested ($p = .01$ in all cases), providing support for the increment in fit at each step.
The smallest profiles in solutions with more than three profiles comprised 8--9\% of the sample.
Balancing statistical fit, classification precision, and interpretability, we retained the three-profile solution as an interpretable summary (\glsentryshort{BIC} $= 9{,}171.34$; \glsentryshort{AIC} $= 9{,}084.72$; Entropy $= 0.90$).

\begin{table}[t]
\caption{Fit statistics for latent profile analysis models (2--7 profiles).
The three-profile solution (marked *) was selected.}
  \label{tab:fit}
  \begin{tabular}{lrrccrr}
    \toprule
    \textbf{Profiles} & \textbf{\glsentryshort{AIC}} & \textbf{\glsentryshort{BIC}} & \textbf{Entropy} & \textbf{\glsentryshort{BLRT} $p$} & \textbf{$n_{\min}$} & \textbf{$n_{\max}$}\\
    \midrule
    2   & 9{,}729.99 & 9{,}792.55 & 0.90 & .01 & 0.21 & 0.79\\
    3*  & 9{,}084.72 & 9{,}171.34 & 0.90 & .01 & 0.11 & 0.69\\
    4   & 8{,}954.13 & 9{,}064.81 & 0.88 & .01 & 0.09 & 0.46\\
    5   & 8{,}352.25 & 8{,}487.00 & 0.89 & .01 & 0.08 & 0.42\\
    6   & 8{,}137.75 & 8{,}296.56 & 0.88 & .01 & 0.09 & 0.39\\
    7   & 7{,}878.12 & 8{,}060.99 & 0.88 & .01 & 0.08 & 0.31\\
    \bottomrule
  \end{tabular}
\Description{Table of fit statistics for two through seven profile solutions, showing \glsentryshort{AIC}, \glsentryshort{BIC}, entropy, bootstrap likelihood ratio test p-values, and the proportion of the sample in the smallest and largest profiles.
\glsentryshort{BIC} decreases across the tested solutions.}
\end{table}

\subsection{Profile characterization}
\begin{figure*}[t]
  \centering
  \includegraphics[width=0.82\textwidth]{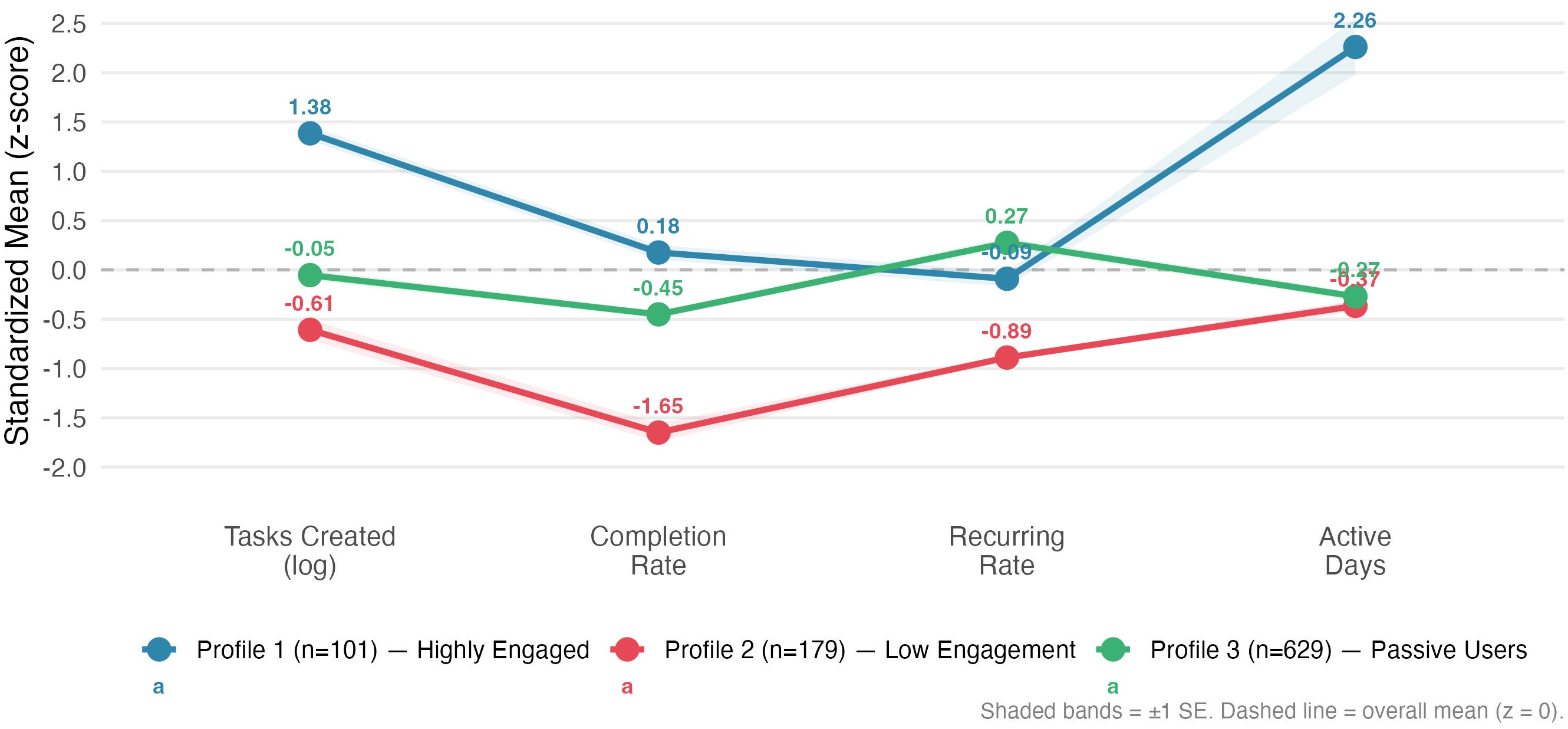}
\caption{Behavioral profiles from the latent profile analysis.
Standardized (z-scored) means on the four core behavioral indicators for the Highly Engaged, Low Engagement, and Passive Users profiles.}
\Description{A line or bar plot comparing three user profiles across four standardized behavioral indicators: tasks created, completion rate, recurring task rate, and active days.
The Highly Engaged profile is far above average on active days and tasks created.}
  \label{fig:profiles}
\end{figure*}

The three-profile solution yielded well-separated, theoretically interpretable profiles.
Standardized means for each behavioral indicator by profile are summarized in Table~\ref{tab:means} and visualized in Figure~\ref{fig:profiles}.

\emph{Profile 1 (Highly Engaged; $n = 101$, 11.1\%)} was characterized by markedly above-average task creation ($z = 1.38$), high task completion rate ($z = 0.18$), a relatively low rate of recurring tasks ($z = -0.09$), and substantially more active days than any other profile ($z = 2.26$).
This profile describes users who engaged with the app intensively and consistently across the observation period, creating many unique tasks on a high number of calendar days.

\emph{Profile 2 (Low Engagement; $n = 179$, 19.7\%)} was distinguished by below-average task creation ($z = -0.61$), markedly low completion rates ($z = -1.65$), a pronounced tendency to leave tasks unfinished, and below-average use of recurring tasks ($z = -0.89$) and active days ($z = -0.37$).
This profile captures users who attempted to use the app but completed relatively few of the tasks they created, suggesting a planning--execution gap.

\emph{Profile 3 (Passive Users; $n = 629$, 69.2\%)} represented the modal pattern of app use.
These users showed near-average task creation ($z = -0.05$) and a moderate above-average tendency to use recurring tasks ($z = 0.27$), but below-average completion rates ($z = -0.45$) and fewer active days ($z = -0.27$).
This profile describes users who set up recurring tasks but engaged with the app sporadically, completing tasks at a rate below the overall sample mean.

\begin{table}[t]
\caption{Standardized profile means (z-scores) on behavioral indicators from the latent profile analysis.
Standard errors in parentheses.}
  \label{tab:means}
  \begin{tabular}{lccc}
    \toprule
    \textbf{Indicator} & \textbf{P1: Highly} & \textbf{P2: Low} & \textbf{P3: Passive}\\
     & \textbf{Engaged} & \textbf{Engagement} & \textbf{Users}\\
     & \textbf{($n=101$)} & \textbf{($n=179$)} & \textbf{($n=629$)}\\
    \midrule
    Tasks created (log)  & 1.38 (0.08)  & $-0.61$ (0.10) & $-0.05$ (0.06)\\
    Completion rate      & 0.18 (0.07)  & $-1.65$ (0.09) & $-0.45$ (0.03)\\
    Recurring task rate  & $-0.09$ (0.08) & $-0.89$ (0.03) & 0.27 (0.05)\\
    Active days          & 2.26 (0.28)  & $-0.37$ (0.05) & $-0.27$ (0.04)\\
    \bottomrule
  \end{tabular}
  \Description{Table of standardized means and standard errors for each of the four behavioral indicators across the three profiles.}
\end{table}

\subsection{Profile differences in subjective outcomes}
\begin{figure}[t]
  \centering
  \includegraphics[width=\columnwidth]{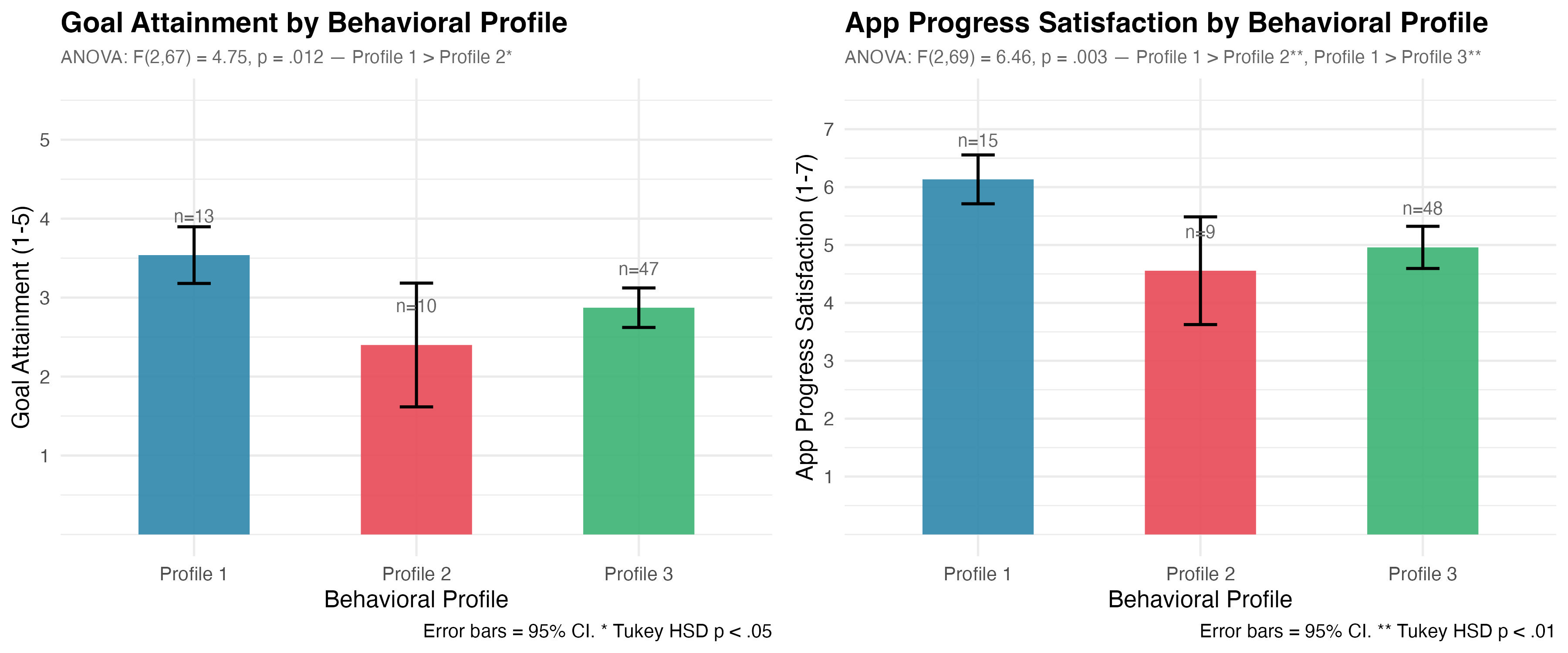}
  \caption{Goal attainment (left) and app progress satisfaction (right) by behavioral profile.}
  \Description{Two grouped bar charts showing mean goal attainment and mean app progress satisfaction for the three profiles, with the Highly Engaged profile highest on both.}
  \label{fig:outcomes}
\end{figure}

Of the 904 matched users with both behavioral and pre-survey data, only between 70 and 74 users had complete data across the five outcome variables (7.7--8.2\%), limiting the power of outcome analyses. Sample sizes for each outcome were as follows: goal attainment (n = 70), app progress satisfaction (n = 72), productivity happiness (n = 71), unproductiveness problematicness (n = 74), post productivity problems (n = 72). Nevertheless, significant profile differences emerged on two of the five outcomes examined; all five are summarized in Table~\ref{tab:anova} and the two significant effects are plotted in Figure~\ref{fig:outcomes}.
We regard these outcome analyses as preliminary and secondary to the lifecycle and behavioral results above, given the low follow-up rate.

\subsubsection{Goal attainment}
A one-way \gls{ANOVA} revealed a significant effect of behavioral profile on follow-up goal attainment, $F(2, 67) = 4.75$, $p = .012$.
Tukey \glsentryshort{HSD} post-hoc comparisons indicated that Highly Engaged users (Profile 1) reported significantly higher goal attainment than Low Engagement users (Profile 2; mean difference $= 1.14$, 95\% \gls{CI} $[0.23, 2.05]$, $p_{\text{adj}} = .011$).
The comparison between Profile 1 and Passive Users (Profile 3) approached but did not reach significance (mean difference $= 0.67$, 95\% \gls{CI} $[-0.01, 1.35]$, $p_{\text{adj}} = .056$).
Profiles 2 and 3 did not differ significantly (mean difference $= -0.47$, 95\% \gls{CI} $[-1.30, 0.35]$, $p_{\text{adj}} = .366$).

\subsubsection{App progress satisfaction}
\begin{figure}[t]
  \centering
  \includegraphics[width=\columnwidth]{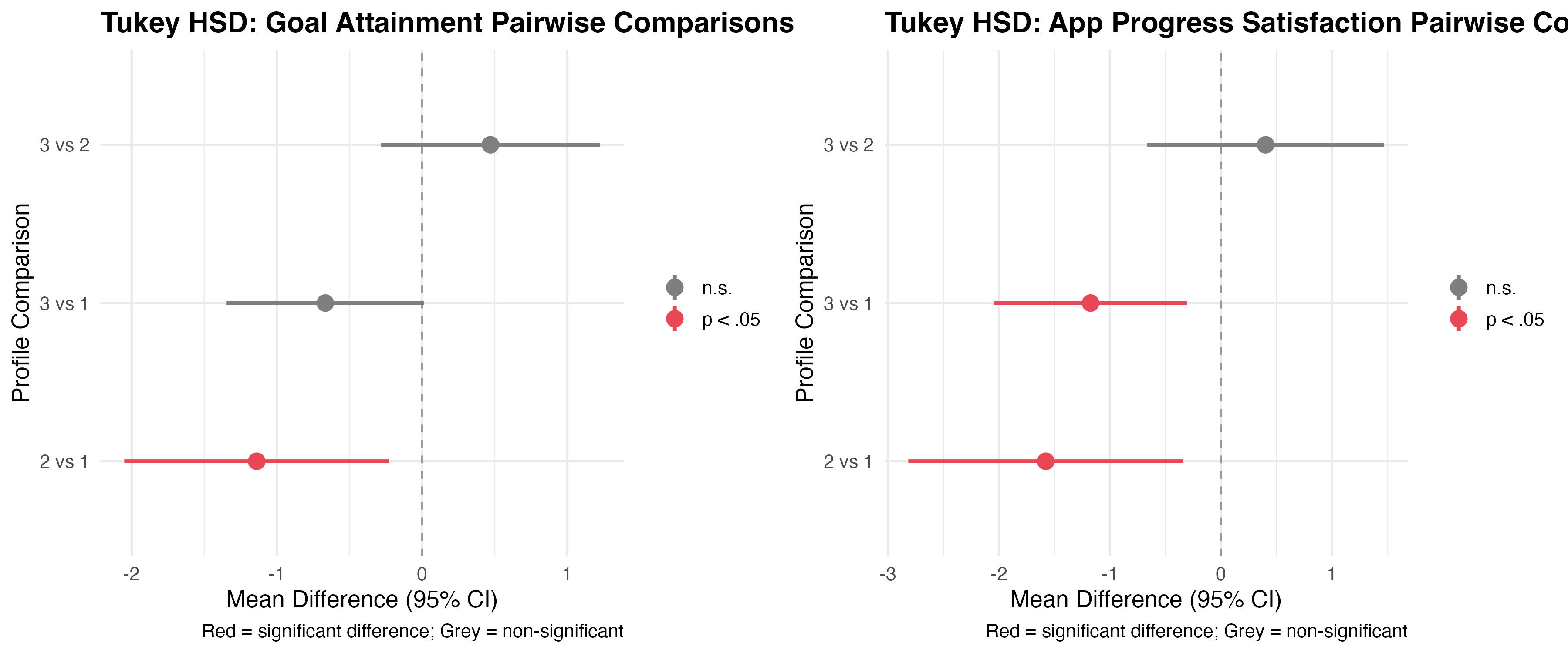}
  \caption{Tukey \glsentryshort{HSD} pairwise comparisons for goal attainment (left) and app progress satisfaction (right).}
  \Description{Forest-plot style visualization of pairwise mean differences with confidence intervals for the three profile comparisons on the two significant outcomes.}
  \label{fig:tukey}
  \vspace{2pt}
  \parbox{\columnwidth}{\footnotesize \textit{Note.} Red indicates $p < .05$; grey indicates non-significant.}
\end{figure}

A one-way \gls{ANOVA} also revealed a significant effect of behavioral profile on follow-up satisfaction with app progress, $F(2, 69) = 6.46$, $p = .003$.
Tukey \glsentryshort{HSD} comparisons showed that Highly Engaged users reported significantly higher progress satisfaction than both Low Engagement users (mean difference $= 1.58$, 95\% \gls{CI} $[0.34, 2.82]$, $p_{\text{adj}} = .008$) and Passive Users (mean difference $= 1.18$, 95\% \gls{CI} $[0.31, 2.04]$, $p_{\text{adj}} = .005$).
Profiles 2 and 3 did not differ significantly (mean difference $= -0.40$, 95\% \gls{CI} $[-1.67, 0.86]$, $p_{\text{adj}} = .727$).
Figure~\ref{fig:tukey} shows all pairwise comparisons with their confidence intervals.

Three additional outcomes showed no significant profile differences.
The pre-to-post difference in productivity happiness did not differ across profiles, $F(2, 68) = 1.72$, $p = .188$.
The pre-to-post change in perceived unproductiveness problematic-ness was likewise not significantly associated with profile membership, $F(2, 71) = 1.27$, $p = .288$.
Post-intervention perceived productivity problems also did not differ across profiles, $F(2, 69) = 1.35$, $p = .266$.
The small, selected follow-up sample limits the precision and generalizability of these estimates; nonsignificant results do not establish equivalence across profiles.

\begin{table}[t]
  \caption{One-way \gls{ANOVA} results for behavioral-profile differences in follow-up outcomes. $\Delta$ = post minus pre difference score.}
  \label{tab:anova}
  \begin{tabular}{lcccl}
    \toprule
    \textbf{Outcome} & \textbf{df} & \textbf{$F$} & \textbf{$p$} & \textbf{Tukey sig.}\\
    \midrule
    Goal attainment            & 2, 67 & 4.75 & .012*  & P1 $>$ P2\\
    App progress satisfaction  & 2, 69 & 6.46 & .003** & P1 $>$ P2, P1 $>$ P3\\
    Productivity happiness ($\Delta$) & 2, 68 & 1.72 & .188 & ---\\
    Unproductiveness ($\Delta$)       & 2, 71 & 1.27 & .288 & ---\\
    Post productivity problems        & 2, 69 & 1.35 & .266 & ---\\
    \bottomrule
  \end{tabular}
\Description{\gls{ANOVA} table listing degrees of freedom, F values, p values, and significant Tukey comparisons for five follow-up outcomes.
Only goal attainment and app progress satisfaction are significant.}
\end{table}

\subsection{Attrition analyses}

\subsubsection{Profile differences in post-survey completion}
\begin{figure}[t]
  \centering
  \includegraphics[width=\columnwidth]{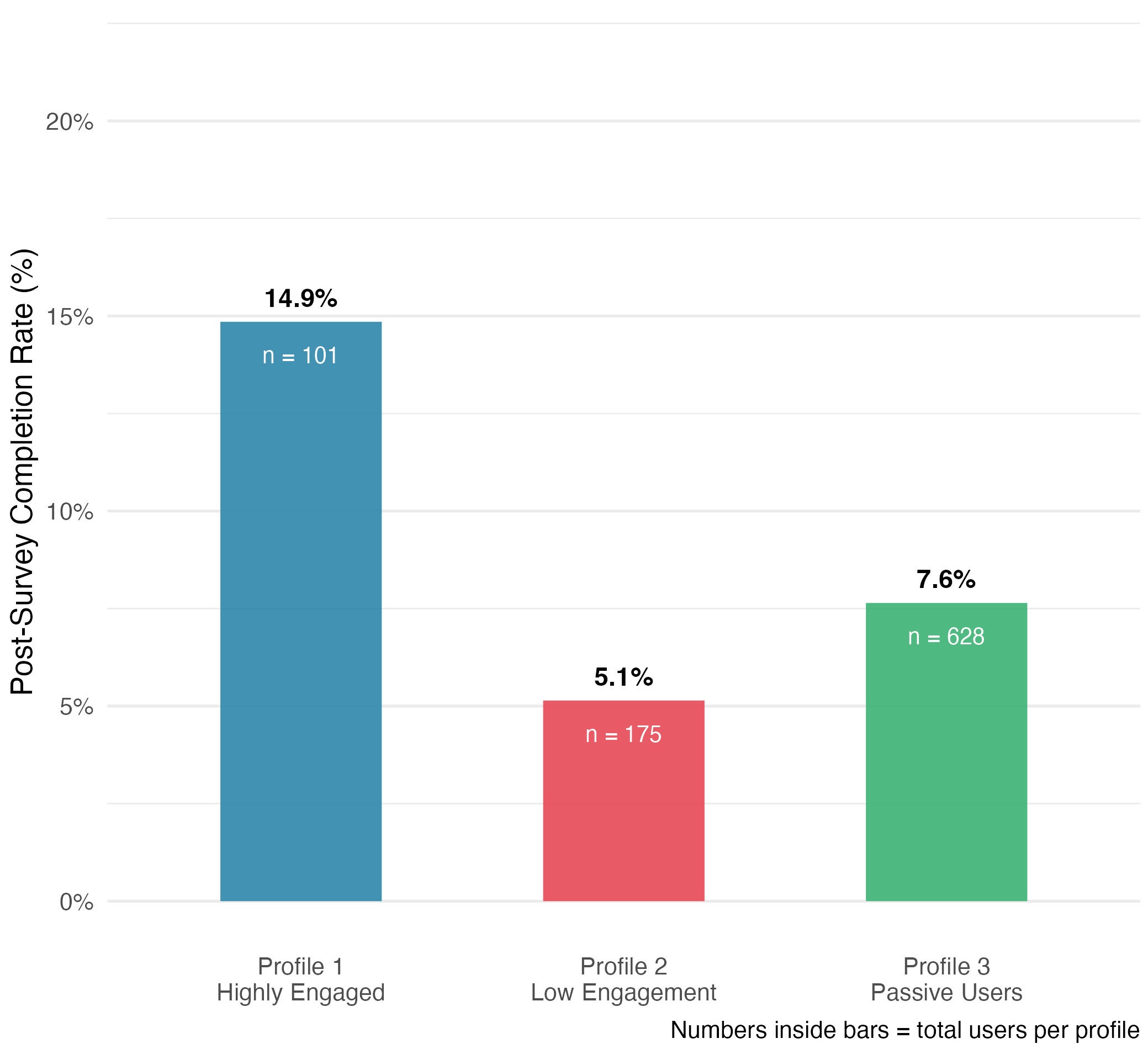}
  \caption{Post-survey completion rate by behavioral profile.}
  \Description{Bar chart of post-survey completion percentages for the three profiles: Highly Engaged highest at about 15 percent, Passive Users intermediate, Low Engagement lowest at about 5 percent.}
  \label{fig:retention}
\end{figure}

Of the 909 users included in the latent profile analysis, 904 were matched with survey records; the remaining 5 had no corresponding survey data and were excluded from survey-linked attrition analyses only. Post-survey completion rates were then examined among the 904 matched users. Post-survey completion rates differed significantly across behavioral profiles, $\chi^2(2) = 8.52$, $p = .014$ (see Figure~\ref{fig:retention}).
Profile 1 (Highly Engaged) showed the highest post-survey response rate, with 14.9\% of users completing the post-survey (15 of 101).
Profile 3 (Passive Users) showed intermediate post-survey response at 7.6\% (48 of 628).
Profile 2 (Low Engagement) had the lowest post-survey response, with only 5.1\% completing the post-survey (9 of 175).
Highly Engaged users were thus nearly three times as likely as Low Engagement users to complete the post-survey.

\subsubsection{Behavioral correlates of post-survey completion}
\begin{figure}[t]
  \centering
  \includegraphics[width=0.72\columnwidth]{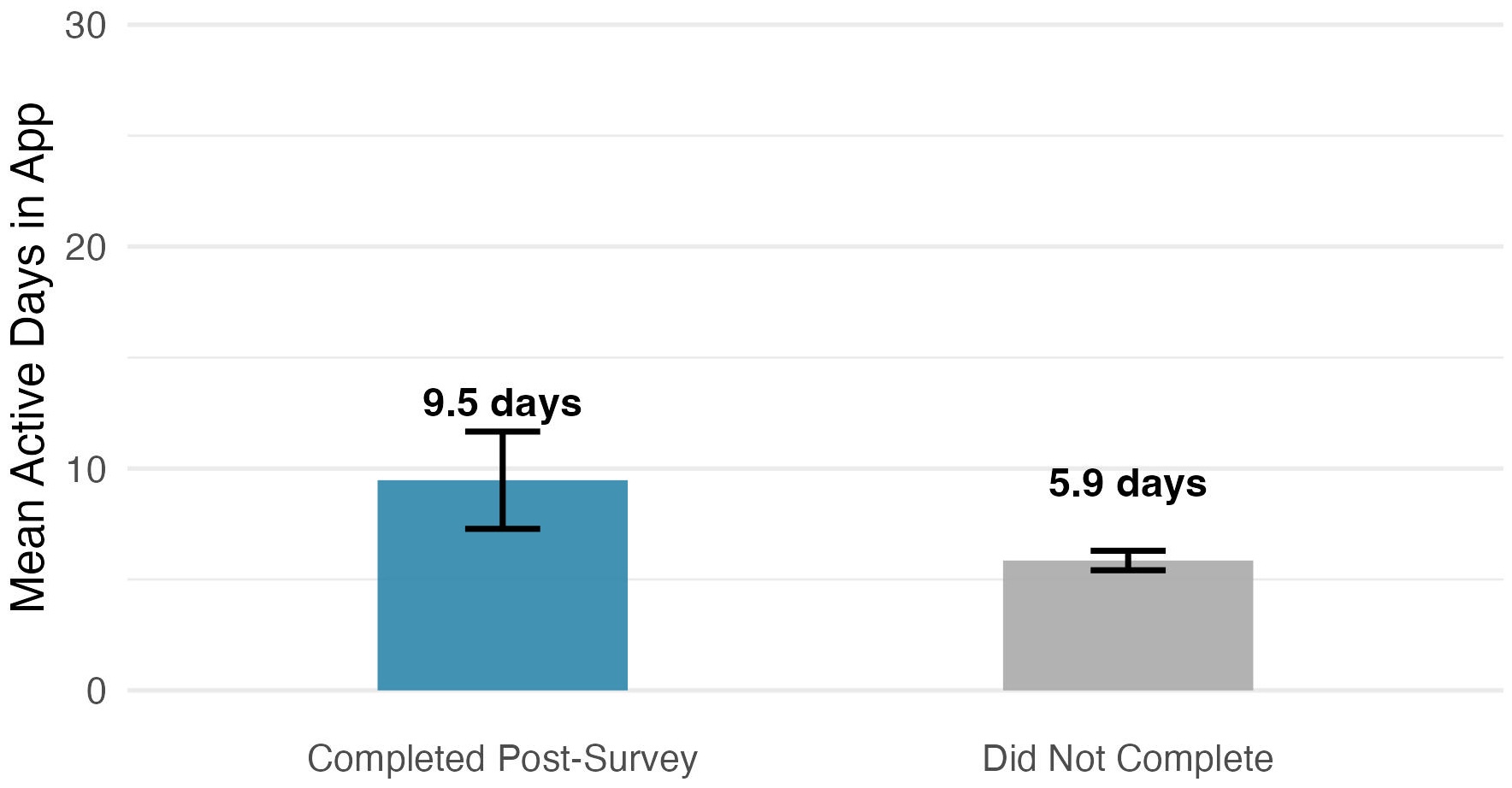}
\caption{Mean active days by post-survey completion status.
Error bars = 95\% \glsentryshort{CI}.}
  \Description{Bar chart comparing mean number of active days for users who completed versus did not complete the post-survey, with completers showing more active days.}
  \label{fig:activedays}
  \vspace{2pt}
  \parbox{\columnwidth}{\footnotesize \textit{Note.} Active days was the only significant predictor of post-survey completion ($\glsentryshort{OR} = 1.05$, $p = .048$).}
\end{figure}

\begin{figure}[t]
  \centering
  \includegraphics[width=0.78\columnwidth]{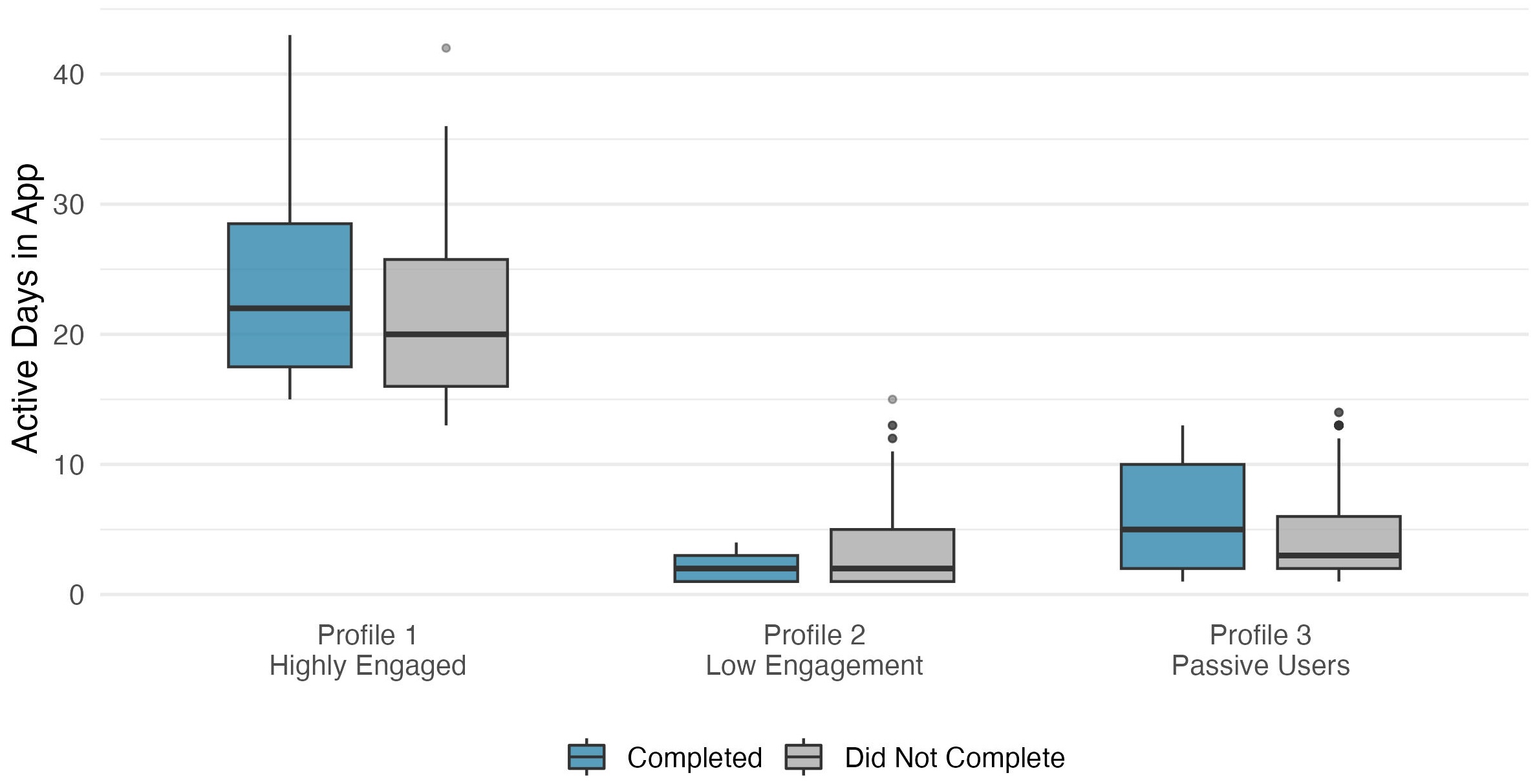}
  \caption{Active days by behavioral profile and post-survey completion status.}
  \Description{Grouped bar chart of mean active days split by the three profiles and by post-survey completion status.}
  \label{fig:activedaysprofile}
\end{figure}

To identify which specific behavioral variables predicted post-survey completion, we fit a logistic regression model with completion status as the outcome and the four behavioral indicators (completion rate, active days, tasks created, recurring-task rate) as predictors (Table~\ref{tab:logit}).
Of the four, only active days significantly predicted post-survey completion ($b = 0.048$, $\glsentryshort{SE} = 0.024$, $z = 1.98$, $p = .048$, $\glsentryshort{OR} = 1.05$, 95\% \gls{CI} $[1.001, 1.102]$).
Completion rate ($\glsentryshort{OR} = 1.27$, $p = .669$), tasks created ($\glsentryshort{OR} = 1.001$, $p = .644$), and recurring-task rate ($\glsentryshort{OR} = 1.28$, $p = .630$) were not significantly associated with post-survey completion in this model.
Figure~\ref{fig:activedays} shows the difference in mean active days between users who did and did not return for the post-survey, and Figure~\ref{fig:activedaysprofile} breaks that difference down by profile.

We then examined whether behavioral-profile membership contributed incremental predictive power beyond the behavioral indicators by adding profile as a factor to the logistic regression.
A likelihood-ratio test comparing the two models was non-significant ($\Delta\text{deviance} = 2.13$, df $= 2$, $p = .344$), providing no evidence that adding profile membership improved model fit beyond the four behavioral indicators.

\begin{table}[t]
\caption{Logistic regression predicting post-survey completion from behavioral indicators.
Outcome: post-survey completion (1 = completed, 0 = not). $\glsentryshort{OR}$ = odds ratio; $N = 904$.}
  \label{tab:logit}
  \begin{tabular}{lrrrrr}
    \toprule
    \textbf{Predictor} & \textbf{$b$} & \textbf{\glsentryshort{SE}} & \textbf{$z$} & \textbf{$p$} & \textbf{$\glsentryshort{OR}$}\\
    \midrule
    Intercept          & $-3.12$ & 0.39  & $-8.10$ & $<.001$ & 0.04\\
    Completion rate    & 0.24   & 0.56  & 0.43   & .669    & 1.27\\
    Active days        & 0.05   & 0.02  & 1.98   & .048*   & 1.05\\
    Tasks created      & 0.001  & 0.002 & 0.46   & .644    & 1.001\\
    Recurring-task rate & 0.24  & 0.51  & 0.48   & .630    & 1.28\\
    \bottomrule
  \end{tabular}
\Description{Logistic regression coefficient table with b, standard error, z, p, and odds ratio for the intercept and four behavioral predictors.
Only active days is significant.}
\end{table}

\section{Discussion}

\subsection{Behavioral profiles and their interpretation}
The three-profile solution provides a substantively coherent account of how new users interact with a digital planner.
The Highly Engaged profile (11.1\% of users) resembles what \citet{reijula2022selfnudging} describe as successful self-nudgers: individuals who actively architect their own behavioral environment using the available tool.
The defining feature of this profile was not merely creating many tasks but doing so across a high number of active days, reflecting consistent behavioral integration of the app into daily routines.

The Low Engagement profile (19.7\%) is particularly theoretically interesting.
These users did not simply use the app rarely; rather, they created tasks at a below-average rate and left the majority of created tasks uncompleted.
This pattern is consistent with a planning--execution gap~\cite{gollwitzer2006implementation}, wherein users intend to be productive but fail to follow through on created plans.
Such a pattern may reflect low implementation-intention strength, there also remain questions regarding one's psychological and contextual factors such as impulsivity, or motivational deficits that the app's scaffolding alone may be insufficient to overcome.

The Passive Users profile (69.2\%) was the modal type and is arguably the most representative of how digital productivity tools are actually used in naturalistic settings.
These users showed average to slightly below-average completion rates and a tendency to set up recurring tasks (perhaps representing aspiration toward routine), but relatively inconsistent day-to-day engagement.
This profile may reflect the common experience of downloading a productivity app with good intentions but failing to incorporate it into a stable behavioral routine, a pattern consistent with the documented tendency for app usage to decline rapidly after installation~\cite{haliburton2024understanding}.

\subsection{Profiles and self-reported outcomes}
The profiles differed not only in behavior but in self-reported outcomes, although these analyses were sharply constrained by low post-survey response: only 70 to 74 of the 904 matched users (7.7--8.2\%), depending on the outcome, had complete follow-up data.
Within that subsample, Highly Engaged users reported higher goal attainment than Low Engagement users, and higher satisfaction with their app progress than both other profiles.
In other words, the users who integrated the tool most fully into their daily routines were also the ones who perceived themselves as making the most progress.
This convergence of behavioral engagement and subjective benefit is encouraging, but it must be read cautiously.
The follow-up sample is small (7.9\% of matched users), and response rates differ by profile.
We therefore treat the outcome findings as exploratory; the lifecycle and behavioral analyses remain the primary contribution.
Three of the five outcomes showed no profile differences, and given the small and non-representative follow-up sample, the estimates do not establish either equivalence or a difference across profiles; the significant differences also require replication.
The direction of any relationship is also ambiguous: engagement may produce benefit, perceived benefit may sustain engagement, or a third factor such as trait conscientiousness or baseline motivation may drive both.

\subsection{From plan-and-execute to a lifecycle account}
The plan-and-execute assumption treats task creation as the psychologically decisive act and completion as routine follow-through.
The task-level data are difficult to reconcile with that view.
Only 26.4\% of created tasks travelled the direct path from creation to completion; 32.0\% were abandoned outright, and about half had no recorded completion within six weeks.
Nor was the plan followed as written even when it was followed: 77.6\% of timed tasks were marked complete later than their intended time, by a median of about an hour.
The typical fate of a task in these data is therefore not the tidy movement from creation to completion that a ledger-of-intentions model presumes, but a plan under continuous renegotiation.

Two further observations sharpen the point.
First, deferral and abandonment behave as distinct processes rather than as degrees of a single failure.
Reschedules were, if anything, associated with marginally higher odds of eventual completion, while abandonment accrued quietly and had largely run its course by day ten.
A framework whose only outcomes are executed and not executed has nowhere to put a task that is alive but moving; yet on these data that is a common state, and arguably the characteristic one.
Second, the same asymmetry appears at the person level.
Reported standardized completion rates differed across profiles, and for the Low Engagement profile the divergence between what was created and what was completed was pronounced (completion $z = -1.65$).
Tellingly, even the users who created the most tasks did not simply execute them all: the advantage of the Highly Engaged profile lay less in flawless completion ($z = 0.18$) than in sustained, distributed engagement across many days ($z = 2.26$).
Observed as behavior rather than as recalled intention, planning looks like a process in which intentions are routinely formed and then partially, belatedly, or never carried out, precisely the shift of emphasis, from the moment of intention formation to what happens to intentions afterward, that motivates a lifecycle view.

\subsection{Temporal dynamics and everyday planning}
Planning had a discernible rhythm.
Creation peaked in the evening and early in the week, completion at midday and midweek, so intentions were characteristically formed at one point in the cycle and discharged at another; the gap between them is where the lifecycle happens.
That gap was also directional.
Timed tasks were overwhelmingly completed later than intended rather than earlier (77.6\%), which suggests that the time a user attaches to a task functions less as a prediction than as an aspiration, and that slippage is the norm rather than the exception.
The competing-risks estimates give this a horizon: completion, when it came, came quickly (median half a day), while abandonment accumulated over roughly ten days and then stopped.
These estimates describe the observed window; they do not establish a deadline after which tasks cannot be completed.

The most diagnostic behavioral dimension in our data was neither the number of tasks created nor even the proportion completed, but the temporal breadth of engagement: the number of active days.
Active days was the feature that most sharply separated Highly Engaged users from the rest ($z = 2.26$), and it was the only behavioral indicator that independently predicted whether a user completed the post-survey ($\glsentryshort{OR} = 1.05$, $p = .048$).
Together these point to an association between the number of active days and follow-up survey response.
Returning to the tool across many days did more to distinguish sustained planners than output on any single occasion.
The modal Passive Users profile makes the complementary point: these users set up recurring tasks, an apparent aspiration toward routine ($z = 0.27$), yet engaged only sporadically and completed at a below-average rate, suggesting that features designed to scaffold routine do not, by themselves, generate the day-to-day return that consistent planning requires.
On this account, everyday planning is less a discrete act than an ongoing practice, and the logs capture variation in how regularly users return to it.

\subsection{Implications for self-regulation and the intention--behavior gap}
The Low Engagement profile is, in effect, the intention--behavior gap rendered in behavioral logs: users created tasks, that is, formed intentions, but completed few of them.
This provides naturalistic, non-retrospective corroboration of a phenomenon usually documented through self-report and laboratory tasks~\cite{sheeran2016intention, webb2006does}, and it suggests that the gap is not a uniform population tendency but a concentrated one.
A minority of users largely closed it through sustained engagement, a substantial minority exhibited it acutely, and the majority sat in between with modest under-completion and sporadic use.
Modeling self-regulatory failure as a distribution of engagement types, rather than as a single average propensity, may therefore be more faithful to how procrastination and self-regulation actually vary across people~\cite{steel2007nature}.
The observation that recurring-task use did not protect Passive Users from under-completion is also relevant to implementation-intention theory~\cite{gollwitzer1999implementation, gollwitzer2006implementation}: instantiating a repeating item in a planner is not the same as forming a strong, context-cued if-then intention, and may lack precisely the specificity that makes implementation intentions effective.

\subsection{Implications for the design of planning tools}
If completion rather than creation is the binding constraint, then tools designed primarily as faithful ledgers of intentions are optimizing the wrong stage of the lifecycle.
Several design directions follow.
First, the association between active days and post-survey response motivates testing lightweight daily check-ins and low-friction re-entry.
Such features should be evaluated against task outcomes as well as continued app use.
The profiles reported here were estimated over the full six-week window, so they describe engagement retrospectively and cannot yet be assigned prospectively; whether the same profiles are recoverable from a user's first days is an open question that would require a separate early-period validation.
Second, and subject to that caveat, if profiles can be identified from early behavior, adaptive support could be tested: Low Engagement users, who create tasks but abandon them, might benefit from execution scaffolding such as context-linked reminders, if-then prompts, and task decomposition, rather than further prompts to capture more; Passive Users might benefit from features that convert nominal recurring tasks into genuinely cued routines; and Highly Engaged users may need little support and could even be hindered by added friction.
This aligns with just-in-time adaptive intervention frameworks~\cite{nahumshani2018jitai} and with the notion of self-nudging, in which the tool helps users architect their own choice environment~\cite{reijula2022selfnudging}.
Third, the lower post-survey response among Low Engagement users makes their experiences harder to assess. Evaluations of adaptive support should account for this selective response.

\subsection{Unique insights and limitations}
The central strength of this work is its data.
Naturalistic usage logs capture what users actually did, moment to moment, free of the memory distortion and social-desirability bias that attend retrospective self-report, and at a scale and granularity that laboratory studies cannot match.
Observing planning in the wild also preserves the competing demands, interruptions, and stakes of real life that controlled tasks strip away, so the behavioral profiles reflect planning as it is lived rather than as it is simulated.

These strengths are accompanied by real limitations.
First, the sample is self-selected: people who download and try a commercial planner are already disposed to plan, and we cannot infer the direction of selection bias or generalize these profile proportions to people in general.
Second, the data are ecological but non-experimental, and therefore correlational; the association between engagement and better self-reported outcomes cannot establish that engagement causes benefit, and reverse causation and third variables remain plausible alternatives.
Third, outcome inference is limited by attrition: with only 7.9\% completing the post-survey and survey response itself differing by profile, the outcome analyses are both underpowered and non-randomly missing, so both significant and nonsignificant outcome estimates should be treated as preliminary.
Fourth, the findings come from a single application with its particular feature set, and the behavioral indicators are proxies whose meaning depends on user habits.
A task marked complete depends on users bothering to mark it, and some completed work is surely never recorded, so ``abandonment'' as measured here conflates tasks genuinely dropped with tasks done but never ticked off, a distinction the logs cannot make and one that would inflate our abandonment estimate.
The same caution applies to rescheduling, whose extreme upper tail is more plausibly an artifact of automatically rolled-forward recurring items than a record of deliberate re-planning, and to the six-week window, which right-censors long-horizon tasks and cannot speak to planning over months.

Finally, the lifecycle and profile analyses run in parallel rather than being joined.
The profiles are estimated from aggregate person-level indicators, so we can say that Highly Engaged users returned on more days, but not yet whether their individual tasks were rescheduled more often, abandoned sooner, or completed closer to their intended time.
Linking the two levels, modeling task-level event histories nested within users, is the natural next step and would let the lifecycle claims be tested against individual differences rather than reported alongside them.

\subsection{Future directions}
Several extensions follow directly.
The most immediate is to join the two levels of analysis, modeling task-level event histories as nested within users so that the lifecycle quantities reported here, time-to-completion, reschedule counts, time-to-abandonment, can be estimated as person-varying rather than pooled, and so that the departure from plan-and-execute can be tested against individual differences rather than reported alongside them.
A related step is to characterize what people plan, not only how.
Task titles were outside the scope of the present analysis set, but a collection designed for the purpose, with consent and privacy safeguards specific to task content, could turn them into a taxonomy of everyday intentions, and answer whether work, health, and domestic tasks have systematically different lifecycles.
A second direction is to link the behavioral profiles to the survey measures of personality and cognition, testing whether Low Engagement tracks lower conscientiousness and higher impulsivity as self-regulation theory would predict.
A third is to strengthen outcome inference with designs that reduce attrition bias, incentivized or in-app micro-surveys and ecological momentary assessment, and, where feasible, with randomized manipulations of features (for example, A/B tests of execution scaffolding aimed at Low Engagement users) that would move the account from correlational profiles toward causal claims.
Profile-aware adaptive support could likewise be tested prospectively, examining whether profiles detectable from early behavior can be used to route users to the support they need.
Finally, replication across different applications and populations will be necessary to establish which profiles are artifacts of a particular tool and which reflect more general regularities in how people plan.

\section{Conclusion}
The task histories show how users revise schedules and record outcomes over time.
Most created tasks did not move directly to recorded completion, and users varied in how often they returned to the app.
More active days were associated with completing the follow-up survey.
A lifecycle account makes revisions, delays, and unresolved tasks visible alongside completion, providing a basis for evaluating support throughout the planning process.

\begin{acks}

DG's work is funded by the Huo Family Foundation and Stanford's Center for Digital Health.
DG has ongoing research projects in collaboration with the Structured app.
PS's work is funded by Stanford's Center for Digital Health.
\end{acks}

\section*{Contribution statement}
Conceptualization: SH, IZ, LK, DG.
Data curation: SH, LM.
Formal analysis: SH.
Investigation: LM, DG.
Methodology: SH, DG.
Project administration: DG.
Resources: LM, DG.
Software: SH.
Supervision: DG.
Visualization: SH.
Writing (original draft): SH, DG.
Writing (review \& editing): SH, IZ, LK, LM, PS, DG.

\section*{Use of generative \glsentryshort{AI}}
Generative \gls{AI} assistants (Anthropic's Claude and OpenAI's Codex) were used to draft and revise portions of this manuscript, including the abstract, parts of the results and discussion prose, and the figure descriptions.
They were also used to check the manuscript for internal consistency.
All study design, data collection, and analysis were carried out by the authors without generative \gls{AI}.
The authors reviewed and verified all content, including all reported statistics, and take full responsibility for the manuscript.

\bibliographystyle{ACM-Reference-Format}
\bibliography{references}

\end{document}